\documentclass[printer]{aa}
\usepackage{graphicx}
\usepackage{txfonts}
\graphicspath{ {plots/} }
\usepackage[]{natbib}
\usepackage{amsmath}
\usepackage{amssymb}
\usepackage{soul}
\usepackage[dvipsnames]{xcolor}
\usepackage{tikz}

\bibpunct{(}{)}{;}{a}{}{,}

\titlerunning{}

\def\kms{km~s$^{-1}$}
\def\teff{\textit{T}_{\text{eff}}}

\def\logg{\text{log}(\textit{g})}

\def\ofe{[\text{O}/\text{Fe}]}
\def\feh{[\text{Fe}/\text{H}]}

\def\alphafe{[\alpha/\text{Fe}]}

\begin{document}
\renewcommand{\arraystretch}{1.5}

\title{Impact of $\langle$3D$\rangle$ NLTE on GCE of oxygen\\with the RAdial Velocity Experiment}

\author{G.~Guiglion \inst{1, 2, 3}}

\institute{Zentrum f\"ur Astronomie der Universit\"at Heidelberg, Landessternwarte, K\"onigstuhl 12, 69117 Heidelberg, Germany \\ \email{guiglion@mpia.de}
\and
Max Planck Institute for Astronomy, K\"onigstuhl 17, 69117, Heidelberg, Germany
\and
Leibniz-Institut f{\"u}r Astrophysik Potsdam (AIP), An der Sternwarte 16, 14482 Potsdam, Germany}

\date{Received 19 December 2024 ; accepted 19 May 2025}

\abstract{The stellar chemical abundances, coupled with stellar kinematics
are a unique way to understand the chemo-dynamical processes that occurs to
build the Milky Way and its local volume as we observe today.}
{However, measuring stellar abundances is challenging as one needs to properly 
address the effect of departure from the Local Thermodynamic Equilibrium (LTE), as 
well as the commonly used 1-dimensional model atmosphere. In this work,
we constrain the chemical evolution of [O/Fe] in FG stars of the RAdial
Velocity Experiment (RAVE) with [O/Fe] abundances derived in non-LTE (NLTE)
and with horizontally- temporally -averaged 3D ($\langle$3D$\rangle$) model atmospheres.}
{Using standard spectral fitting method, we determine for the first time
LTE and NLTE abundances of oxygen from the \ion{O}{I} triplet at 8\,446\,\AA~in
turn-off and dwarf stars thanks to intermediate-resolution RAVE spectra,
assuming both 1D and ($\langle$3D$\rangle$) model atmosphere.}
{We find that NLTE effects plays a significant role when determining oxygen
even at a resolution of R= 7500. Typical NLTE-LTE corrections of the order of
-0.12 dex are measured in dwarfs and turn-off stars using 1D MARCS models. In addition,
$\langle$3D$\rangle$ modelling significantly impacts the oxygen abundance measurements.
In contrast to applying $\langle$3D$\rangle$ NLTE abundance corrections or the classical
1D LTE, the full $\langle$3D$\rangle$ NLTE spectral fitting yields improving the precision
of abundances by nearly 10\%. We also show
that the decrease of [O/Fe] in the super-solar metallicity regime is rather characterised
by a flat trend when [O/Fe] is computed in $\langle$3D$\rangle$ NLTE
from full spectral fitting. We attribute this
flattening at super-solar [Fe/H] to the interplay between locally born stars with negative
[O/Fe] and stars migrated from the inner MW regions with super-solar [O/Fe], supporting
the complex chemo-dynamical history of the Solar neighbourhood.}
{Our results are key for understanding the effects of $\langle$3D$\rangle$ and NLTE when
measuring oxygen abundances in intermediate-resolution spectra. NLTE effect should be taken
into account when confronting Galactic chemical evolution models to observations. This
work is a test bed for the spectral analysis of the 4MOST low-resolution
spectra that will share similar properties as RAVE spectra in the red
wavelength domain.}


\titlerunning{oxygen in RAVE}
\authorrunning{G. Guiglion}
\maketitle

\section{Introduction}

Oxygen (together with other $\alpha$ elements such as Ne, Mg, and Ca)
is mainly produced during the hydrostatic burning phases in stars with masses
M$>10$M$_{\bigodot}$. The yields of massive stars are still not yet accurately
constrained and depend mainly on the supernova model (e.g. mass loss and nuclear
reaction rates; \citealt{woosley_1990}). Those elements are diffused into the ISM
via type II SNe explosion. Oxygen is not processed during stellar evolution, hence O
abundances measured in stellar atmospheres is an accurate indicator of enrichment
of the interstellar matter (ISM) with oxygen (see for instance
\citealt{Chiappini2003, Romano2022}).
The Fe peak elements are mainly formed during explosive nucleosynthesis in
type Ia SNe, on long time scale (typically 1-2 Gyr). Hence the ratio between O and
Fe abundance, namely
[O/Fe]\footnote{[O/Fe]=log$\Bigr[\frac{\mathrm{N(O)}}{\mathrm{N(Fe)}}\Bigr]_{\star}$ - log$\Bigr[\frac{\mathrm{N(O)}}{\mathrm{N(Fe)}}\Bigr]_{\bigodot}$} gives us clues on
the respective contributions of type Ia and II SNe to the ISM enrichment.
For instance, large [O/Fe] ratios, typically above 0, indicate that the region where
the stars formed went through high star formation rate, and fast chemical enrichment
(e.g. \citealt{Matteucci1986}). However, measuring oxygen abundances in
stellar atmosphere can be a challenging task. 

Largely used for computing stellar atmospheric parameters and chemical
abundances in large scale surveys, the Local Thermodynamic Equilibrium
(LTE) approximation remains the main pillar of spectral analysis (e.g.
\citealt{nordstrom2004, lee2011, Adibekyan2011, mikolaitis_2014, guiglion2016, guiglion_2018, Stonkute2020, DelgadoMena2021}).
However, the community has put strong efforts to take into account
departures from LTE (e.g.
\citealt{Mishenina2000, Mashonkina2008, Lind2011, Bergemann2013, BergemannNordlander2014, Amarsi2020, Sitnova2021}).
The ultimate goal is to combine NLTE effects with a physically realistic
stellar model atmosphere by transitioning from largely adopted 1-dimensional
geometry and hydrostatic equilibrium to 3-dimensional hydrodynamic
simulations of stellar atmospheres (e.g.
\citealt{Nordlander2017, Wang2024, Lind2024, Storm2023}). NLTE and $\langle$3D$\rangle$
effects on the 6\,300 \& 7\,774\,\AA~ oxygen lines are rather well known, with the
7\,774\,\AA~line mainly sensitive to $\langle$3D$\rangle$ NLTE
\citep[e.g.][]{Caffau2008, Amarsi2019, Bergemann2021}. Understanding such systematics
is fundamental for understanding the chemical abundance patterns measured by
large-scale spectroscopic surveys and constraining Galactic chemical evolution
models and stellar yields.

Among the large scale spectroscopic surveys that helped us understanding
the chemo-dynamical evolution of the Milky Way, the RAdial Velocity
Experiment achieved pioneering work in this domain
(\citealt{steinmetz2006, Steinmetz2020a, Steinmetz2020b}, and references there-in). RAVE targeted more
than half a million stars in the south hemisphere, and the spectrograph
is characterised by an intermediate spectral resolution
$R= \lambda /\delta \lambda \sim7\,500$, with a wavelength
range centered on the \ion{Ca}{II} infrared triplet ($\lambda \in
[8\,420 - 8\,780]$\,\AA). Originally designed as a spectroscopic survey for
radial velocity measurements,  RAVE demonstrated that significant information
is present in such a limited wavelength range. The RAVE consortium determined
atmospheric parameters and chemical abundances of several elements, such as
$\alphafe$, [Mg/Fe], [Fe/H], and [Ni/Fe], using both standard spectroscopy
\citep{boeche2011, kordopatis2013, Steinmetz2020a} and machine--learning methods
\citep{casey2017, guiglion2020}. With such high-quality data, numerous scientific
studies have been achieved, for instance from chemo-dynamics
\citep{boeche2014, Binney2014, minchev_2014, Kordopatis2016, Wojno2018}, to
study of metal-poor stars \citep{matijevic2017}, asteroseismology
\citep{Valentini2017}, open clusters \citep{Conrad2017}, and solar twins
\citep{Jofre2017}.

\begin{figure*}
\centering
\includegraphics[width=\textwidth]{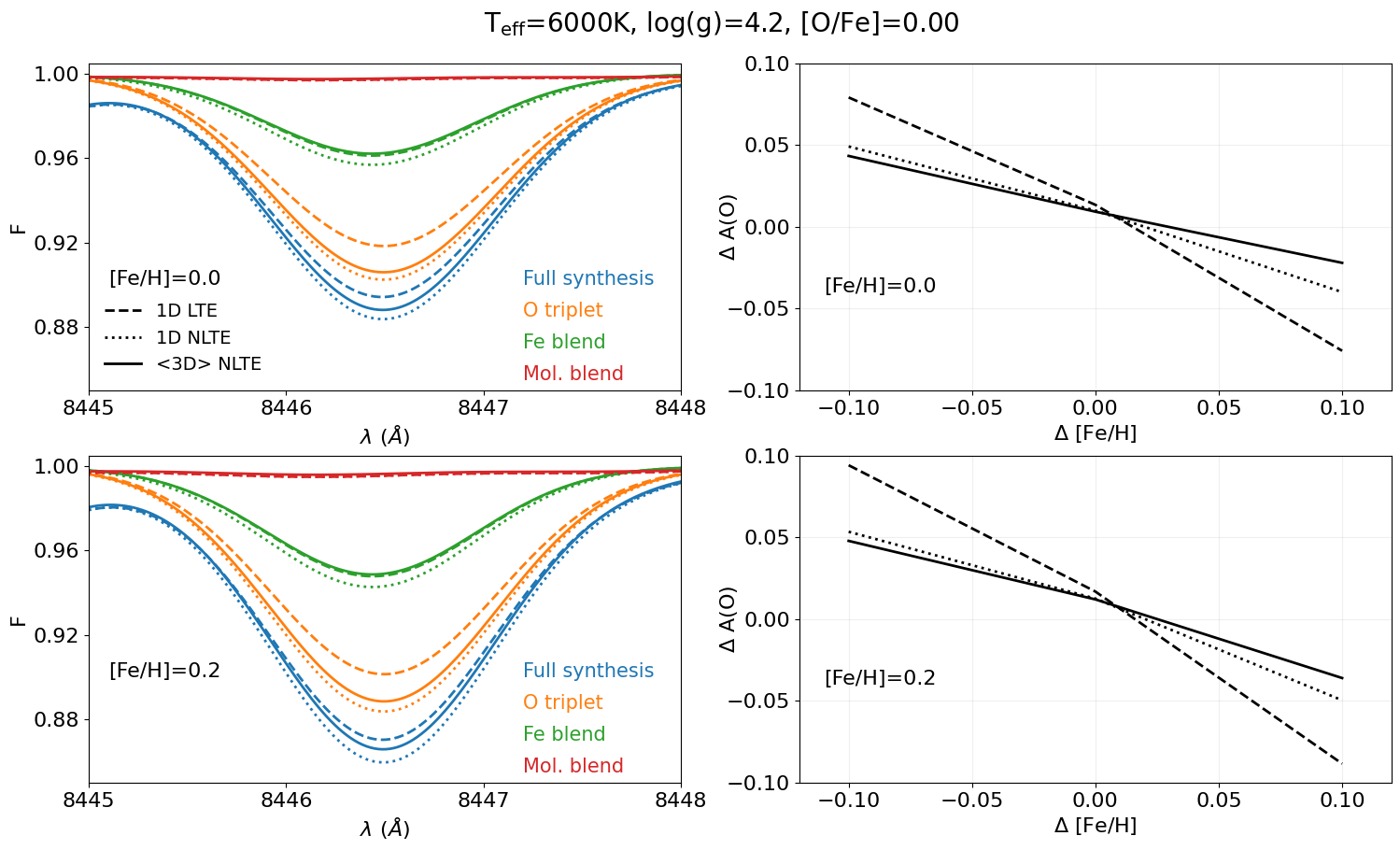}
\caption{Left row: synthetic spectra computed at
$\teff=6\,000\,$K, log(g)=4.2, [O/Fe]=0, in 1D LTE (dashed), 1D NTLE (dotted),
and $\langle$3D$\rangle$ (solid) at RAVE resolution. Full synthesis is shown
in blue, while molecules, O, and Fe blend are shown in red, orange, and green,
respectively. Computation was done at [Fe/H]$=+0.0$ (top), and [Fe/H]$=+0.2$ (bottom).
Right column: sensitivity curves of derived A(O) as a function of a change in
[Fe/H] of $\pm0.1$dex, in 1D LTE, 1D NLTE, and $\langle$3D$\rangle$, at
[Fe/H]$=+0.0$ (top), and [Fe/H]$=+0.2$ (bottom).}
\label{syn_OFe}
\end{figure*} 

Besides the last data release of RAVE been published in 2020,
there is still an opportunity to extract more from such a unique
dataset. In particular, one element remains to be measured in RAVE
spectra using standard spectroscopy analysis, namely oxygen (O),
that can be accessible via the O I triplet lines at 8446\,\AA. One study,
\citet{casey2017}, parametrized oxygen abundances in RAVE spectra using
the machine-learning algorithm Cannon \citep{ness2015}, based on a limited
training sample (less than 1\,400 stars). This study measured oxygen assuming
local thermodynamic equilibrium (LTE). Measuring oxygen at such a resolution
with RAVE spectra is also very relevant for preparing the data analysis of
the upcoming survey 4MOST \citep{4MOST}. Indeed, one of the main
science driver of 4MOST is to target more than 15 million stars in
the Galactic disc and bulge at high- and low-resolution (4MIDABLE-HR,
\citealt{bensby2019}, 4MIDABLE-LR; \citealt{chiappini2019}), Milky Way halo
\citep{Helmi2019}, as well as the Magellanic clouds \citep{Cioni2019}. The
low-resolution optical spectrograph ($\lambda \in [3\,700-9\,500]$\,\AA)
has $R\sim5000-7000$, i.e. very similar to RAVE in the red wavelength domain
and it covers the O I triplet at $7\,774\,$\AA. In the high-resolution spectra,
O abundances in 4MOST will be derived from the [O I] line at 6\,300\,\AA~.
The O I line at 8\,446\,\AA~will present the advantage of naturally having
slightly larger resolution that the two other lines, as well as higher
signal-to-noise ratio (S/N). However, no previous study attempted to
characterise in detail the NLTE and $\langle$3D$\rangle$ effects
on oxygen abundances derived with the triplet at 8\,446\,\AA. The goal of
the present work is to perform the first and comprehensive $\langle$3D$\rangle$
NLTE analysis of oxygen abundances in RAVE spectra with the 8\,446\,\AA~triplet.

In Section~\ref{behaviour_oxygen}, we present an overview
of the oxygen 8\,446\,\AA~triplet region at RAVE spectral resolution. In
Section~\ref{lte_oxygen_sun}, we measure oxygen abundance in the Sun in
1D LTE, and we show how NLTE and horizontally- and temporally-averaged
3D model atmospheres (⟨3D⟩) influence such a measurement. In
Section~\ref{lte_nlte_abundance_from_rave}, we present 1D/$\langle$3D$\rangle$,
LTE and NLTE chemical abundances of oxygen for 8\,018 RAVE dwarfs, while
in Section~\ref{chem_evol_rave}, we perform a chemical evolution analysis of
oxygen in the Milky disc. We present our summary and conclusions in
Section~\ref{conclusiooooonnnn}.

\section{The behaviour of the oxygen triplet at 8\,446\,\AA}\label{behaviour_oxygen}

The oxygen feature at 8\,446\,\AA~is represented by three
strong fine structure \ion{O}{I} transitions at 8\,446.25,
8\,446.36, and 8\,446.76\AA~\citep{Kramida2024}. These transitions
take place between the very high-excitation energy states
($9.52$ eV - $10.99$ eV) between the terms
3s $^{3}$S$^{\mathrm{o}}$ - 3p $^{3}$P, and the total angular
momenta of the upper energy states of $J_{\rm{upper}}=0, 2$,
and 1, respectively for the three transitions.

Due to the limited resolution of RAVE spectra, we expect blends to
influence our capabilities to determine oxygen abundances from such a spectral
feature, especially in the cool temperature regime. To understand the contribution
of the oxygen lines, atomic blends, and possible presence of molecules, we
generated a series of synthetic spectra in 1D LTE, 1D NLTE, and $\langle$3D$\rangle$
NLTE with typical temperature, gravity, and metallicity of stars from our RAVE
sample (see Section~\ref{chem_evol_rave}): $\teff=6\,000\,$K, log(g)=4.2, and
[Fe/H]=0.0 and [Fe/H]=0.2, at solar [O/Fe]. In what follows, we provide
the details on the calculations of synthetic spectra with different physical
models, and explore the influence of blends in the analysis of stellar [O/Fe]
abundance ratios.

\subsection{1D LTE synthetic spectra}\label{1d_lte_syn}

We used the LTE version of the Turbospectrum \citep{plez_2012}
wrapper for synthetic spectra generation, which is part of the spectral
fitting method TSFitPy\footnote{https://github.com/TSFitPy-developers/TSFitPy}.
We adopted 1D hydrostatic MARCS model atmospheres from \citet{gustafsson_2008}.
The Solar abundances are taken from \citet{Magg2022}. We took
advantage of the atomic data from \citet{heiter2021} that compiled high-quality
atomic data for the \emph{Gaia}-ESO survey. The oscillator  strength
of the oxygen triplet lines at 8\,446\AA~ are adopted from \citet{Hibbert1991},
with log(gf)s = -0.463, 0.236, and 0.014, respectively. In our computation,
we also included molecules from \citet{heiter2021}, present in the range
$8\,440-8\,450\,$\AA: FeH, CaH, ZrO, TiO, OH, CN, and CC.

\subsection{1D NLTE synthetic spectra}

It is well known that using LTE approximation can lead to
large biases in the measured chemical abundances (see for
instance \citealt{Hansen2013, Guiglion2024} for barium and strontium,
\citealt{Bergemann2013} for silicon, \citealt{Bergemann2012,
Nordlander2017} for iron and titanium, \citealt{Bergemann2019}
for manganese, and \citet{Wang2024} for lithium).

In the present paper, we quantify the NLTE effect on the
determination of oxygen abundances in RAVE spectra from the
8\,446\AA~triplet. We generated 1D NLTE synthetic spectra using
the NLTE version of Turbospectrum
\footnote{github.com/bertrandplez/Turbospectrum\_NLTE}
\footnote{We note that the NLTE
version of Turbospectrum uses grids of NLTE departure coefficients
to compute NLTE profiles of lines. This is done by correcting
the opacity and the source functions of all lines. These grids
of corrections were computed for the 1D MARCS and averaged 3D
model atmospheres and publicly released as a part of the
original paper on Turbospectrum NLTE v.20 (see \citealt{Gerber2023}
for more details).}
from \citet{Gerber2023}. We adopted the O
model atom of \citet{Bergemann2021}, that was extensively
tested on different solar observations and was used for
constraining the Solar photospheric oxygen abundance (see
also \citealt{Magg2022}). We note that even if [Fe/H] in
LTE is fed to the TSFitPy, the pipeline computes departure
from NLTE for Fe.

\subsection{$\langle$3D$\rangle$ NLTE spectra}\label{syn_3d_nlte}

We computed $\langle$3D$\rangle$ NLTE synthetic
spectra, adopting the spatially- and temporarily-averaged
3D model atmospheres from the STAGGER grid \citep{Magic2013}.
These are full 3D radiation-hydrodynamics simulations of outer
parts of convective envelopes of FGK stars and their averages\footnote{Averages of temperature, density, and electron pressure were carried out on surfaces of equal optical depth (log $\tau_{5000}$).}
in the layers above $\log\,\tau\sim1.5$ can be used as
parameter-free realistic thermodynamic structures, in place of
simplified 1D models in hydrostatic equilibrium, as it has been
demonstrated in a vast body of recent literature
\citep{Bergemann2012, Osorio2015, Bergemann2017, Magg2022, Gerber2023}.

Synthetic spectra are shown in \figurename~\ref{syn_OFe},
at the same spectral resolution as RAVE, i.e. $R\sim7\,500$. In
the left panel, we show spectra computed for a representative star
in our sample ($\teff = 6000$ K, log(g)=4.2) at [Fe/H]$=0.0$ ($\xi=1.184$\,\kms),
while spectra generated with [Fe/H]$=+0.2$ ($\xi=1.195$\,\kms) are presented
in the bottom left corner. We generated series of spectra with only
contribution from molecules (in red), spectra with only contribution
from atomic blends (Fe, in green), spectra with only the oxygen
triplet (orange), and spectra with oxygen triplet plus molecules
and atomic blends (blue). We observe that, independently of [Fe/H],
the molecules do not contribute to the opacity. The atomic blend
is mostly due to Fe I lines, especially two lines at 8446.385 and
8446.575\,\AA~with the lower level excitation energy of 4.988 or
4.913 eV, and log(gf) of -0.920 and -1.440, respectively
\citep{heiter2021}. The strength of the Fe blend is similar
in 1D LTE, and $\langle$3D$\rangle$ NLTE, in both [Fe/H] regimes.

\subsection{Sensitivity of A(O) to [Fe/H]}

We explore how sensitive is the determined A(O) abundance
to the changes in metallicity [Fe/H]. We do this by changing the
input [Fe/H] by $\pm0.1\,$dex (which is typically the uncertainty
on [Fe/H] for the stars in the present paper), at [Fe/H]=+0.0, and
[Fe/H]=+0.2. As in the previous sections, we adopted $\teff=6\,000\,$K,
log(g)=4.2, and [O/Fe]=0.0 as input parameters. We used the code
TSFitPY to perform on-the-fly spectral fitting in a window of
3.5\AA~centered on the oxygen triplet. Results are presented in the
right column of \figurename~\ref{syn_OFe}. In 1D LTE, an uncertainty
of $\pm$0.1dex in [Fe/H] implies a sensitivity of $\pm0.08$dex in
A(O) at [Fe/H]=+0.0, and $\pm0.10$dex in A(O) at [Fe/H]=+0.2 (dashed
lines). Interestingly, the lowest sensitivity of A(O) to [Fe/H] is
achieved in $\langle$3D$\rangle$ NLTE, between 0.03 and 0.05\,dex
(solid line).

\section{Characterizing the NLTE and $\langle$3D$\rangle$ effects on Solar oxygen at RAVE resolution}\label{lte_oxygen_sun}

\subsection{1D LTE}

\begin{figure}
\centering
\includegraphics[width=\columnwidth]{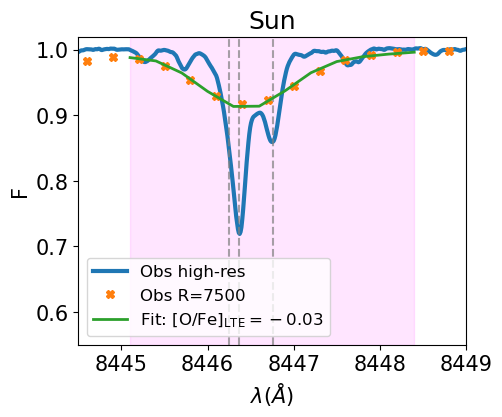}
\caption{High-resolution observation of the Sun from \citet{Neckel1999}
(blue), together with its degraded version to RAVE spectral resolution
(orange). The best fit spectrum, corresponding to $\ofe=-0.03$ (1D LTE),
is shown in green. The spectral fitting window is shown in light purple.}
\label{Sun_Fit}
\end{figure}

We used the ultra-high resolution Solar atlas from \citet{Neckel1999}
that we degraded to RAVE spectral resolution (see
\figurename~\ref{Sun_Fit}). We adopted the standard
solar parameters of $\teff=5\,777$\,K, $\logg=4.44$, and $\feh=0.0$
\citep{Bergemann2012b, Jofre2015}. The abundance computation was done with TSFitPY adopting
1D hydrostatic MARC model atmosphere, with the linelist described
in the previous section. When computing the best fit spectrum, we
naturally included a projected rotation velocity $V\sin i$ of 1.7\kms
\citep{Pavlenko2012}, even if such contribution has a marginal effect
on the quality of the fit at RAVE resolution. 

\begin{figure*}
\centering
\includegraphics[width=\textwidth]{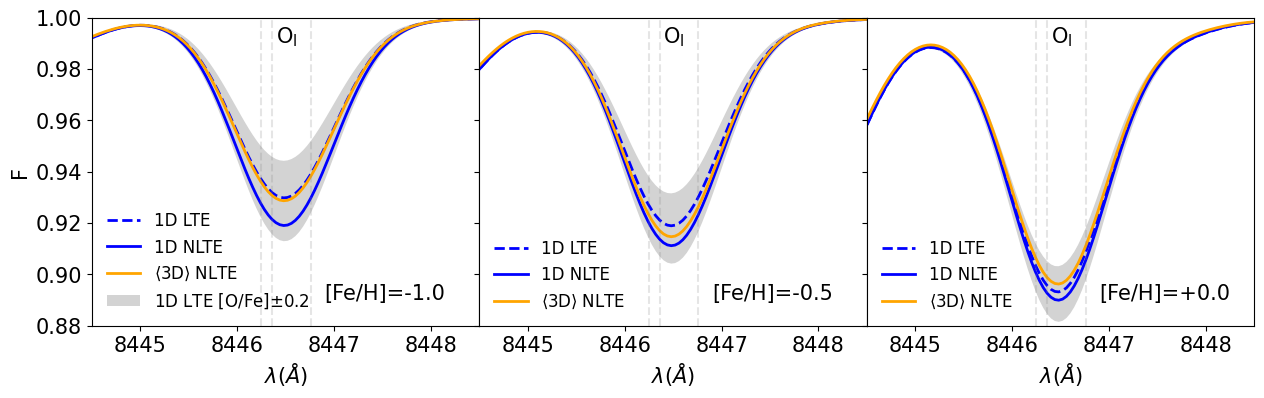}
\caption{Synthetic spectra computed with $\teff=5\,777\,$K,
$\logg=4.44$, and $\ofe=0.0$; LTE and NLTE spectra are shown with
dashed and solid lines, respectively. 1D and $\langle$3D$\rangle$
computations are shown in blue and orange respectively. Left: $\feh=-1$.
Middle: $\feh=-0.5$. Right: $\feh=0.0$. We also add in each panel
synthetic spectra computed at $\ofe=\pm0.2$ in 1D LTE.}
\label{Sun_syn}
\end{figure*} 

To determine the Solar oxygen abundance uncertainty, we propagated
uncertainties on Solar $\teff$, $\logg$, and $\feh$, by performing a
series of six spectral fit by varying one by one the parameter. We
are considering typical uncertainties reported by \citet{guiglion2020}
for solar like stars at RAVE spectral resolution: 70\,K in $\teff$,
0.07 in $\logg$, and 0.04 in $\feh$. The resulting individual
uncertainties when changing a given atmospheric parameter are:
$\sigma_{\mathrm{[O/Fe]}}=\pm0.09$ in $\teff$,
$\sigma_{\mathrm{[O/Fe]}}=\pm0.05$ in $\logg$,
$\sigma_{\mathrm{[O/Fe]}}=\pm0.03$ in $\feh$. Quadratically combining
these three sources of uncertainties gives us
$\ofe_{\mathrm{1D~LTE, \bigodot}}=-0.03\pm0.11$.

\subsection{1D NLTE}\label{1d_nlte}

\begin{figure}
\centering
\includegraphics[width=\columnwidth]{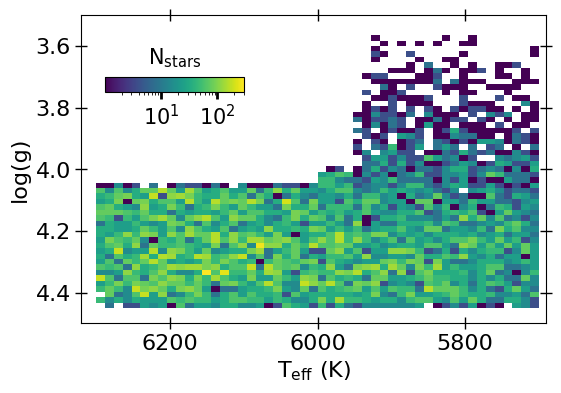}
\caption{Surface gravity vs. effective temperature of the 8\,018 RAVE stars.}
\label{fig:Kiel_diagram}
\end{figure}

We show in \figurename~\ref{Sun_syn} 1D-LTE (blue, dashed) and
1D-NLTE synthetic spectra (blue, solid) at $R=7500$ for
$\feh=[-1.0, -0.5, +0.0]$ and at Solar $\teff$, $\logg$, and $\ofe$.
We clearly see that NLTE affect the 8446\AA~triplet by making the
spectral feature deeper, due to the increase in opacity. It is well
know that the NLTE effects in the triplet lines are mostly driven by
photon losses in the lines (e.g. \citealt{Bergemann2014}); hence the
line source function deviates from the Planck function. We notice an
increasing effect of NLTE when going to lower $\feh$. To guide the eye,
we also plotted 1D-LTE spectra at $\pm0.2$ in $\ofe$. We can see that
the departure from LTE in the line strength is rather large compare to
the change in LTE from $\ofe=0$ to $\ofe=+0.2$, meaning that we expect
NLTE to play a role when determining oxygen abundances at RAVE spectral
resolution.  In the same fashion as in the previous section, we measured
1D-NLTE oxygen abundance in the Solar spectrum at RAVE resolution. We use
the NLTE version of TSFitPy. 

Our resulting solar 1D NLTE O abundance is
$\ofe_{\mathrm{1D~NLTE, \bigodot}}=-0.14\pm0.09$, lower by 0.09 dex compared
to the 1D-LTE abundance. This effect of NLTE can be directly compared with
an independent estimate NLTE abundance correction obtained using MAFAGS-OS
stellar model atmospheres and a different suit of NLTE codes (DETAIL, SIU)
\citep{Bergemann2012, Bergemann2019}. We adopted this estimate from the
\url{nlte.mpia.de} database, finding $\Delta \rm(NLTE-LTE)$ of $-0.06$ for
the $8446$ \AA~lines. We thus conclude that the NLTE correction based on
MARCS$+$MULTI$+$Turbospectrum is in agreement with the estimates based on
MAFAGS-OS$+$DETAIL$+$SIU. The small difference of $0.03$ dex is acceptable
given the substantial differences in the physical and numerical aspects of
these stellar atmospheres and spectral synthesis codes.

\subsection{$\langle$3D$\rangle$ effects in LTE and NLTE}

We aim here at quantifying the effects of LTE and NLTE oxygen
abundances in the Sun when using $\langle$3D$\rangle$ model atmospheres.

Similarly to Section~\ref{1d_nlte}, we computed synthetic spectra at
$R=7500$ for $\feh=[-1.0, -0.5, +0.0]$ and at Solar $\teff$, $\logg$,
and $\ofe$, in $\langle$3D$\rangle$ NLTE. Spectra are plotted in orange
in \figurename~\ref{Sun_syn}. 

In LTE, oxygen feature will be weaker in $\langle$3D$\rangle$ (orange-dashed)
compared to 1D (blue-dashed), independently of [Fe/H]. We observe the same
behaviour in NLTE. It is consistent with the findings in  \citet{Bergemann2021}
and \citet{Magg2022} with oxygen line at 777nm. Using TSFitPy to measure
$\langle$3D$\rangle$ NLTE [O/Fe] in the Sun, we find
$\ofe_{\mathrm{\langle3D\rangle~NLTE, \bigodot}}=-0.01\pm0.09$. We also
computed $\langle$3D$\rangle$ LTE [O/Fe] in the Sun, and found
$\ofe_{\mathrm{\langle3D\rangle~LTE, \bigodot}}=+0.12\pm0.10$, which is higher
compared to the $\langle$3D$\rangle$ NLTE value.

\begin{figure*}
\centering
\includegraphics[width=\textwidth]{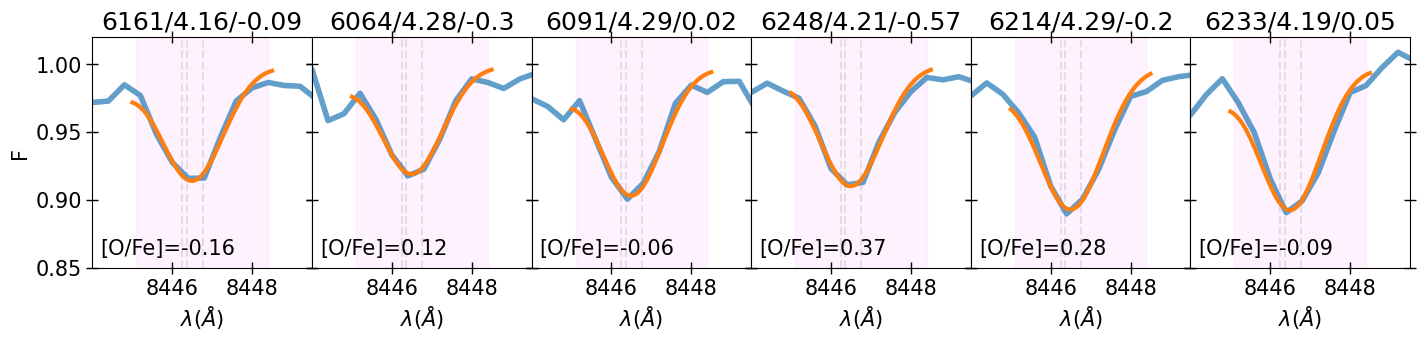}
\caption{Example of spectral fit (orange) around the oxygen triplet at
8\,446\,\AA~(vertical  grey dashed lines) in RAVE spectra (blue). In this
figure, the RAVE spectra are characterised by 50$<$S/N$<$60. The atmospheric
parameters of a given star are indicated at the top of each panel, in the
form $\teff/\logg/\feh$. The fitting window is indicated in light purple.}
\label{spectra_obs}
\end{figure*}

\begin{figure*}
\centering
\includegraphics[width=\textwidth]{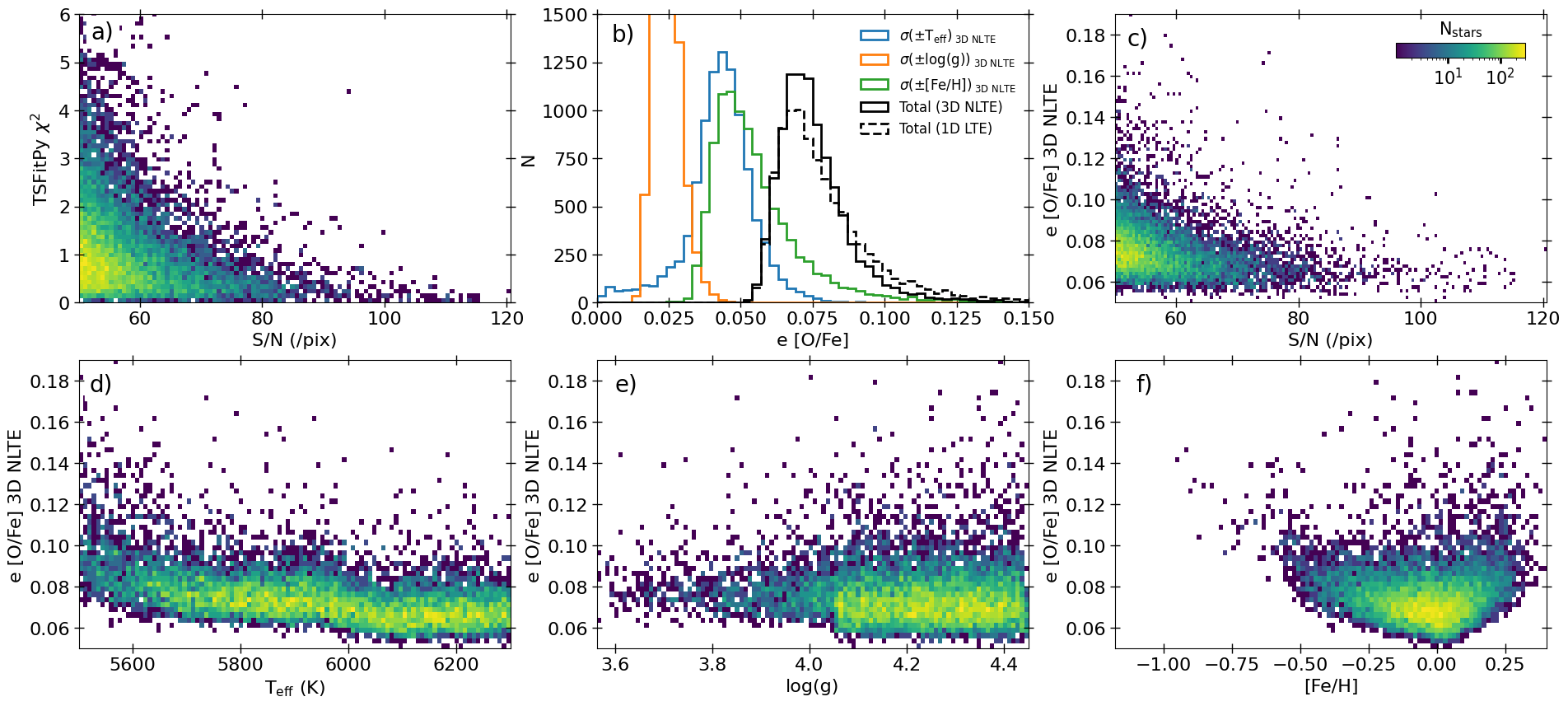}
\caption{a) $\chi^2$ between RAVE observation and best fit spectrum in
$\langle$3D$\rangle$ NLTE around the oxygen triplet at 8\,446\,\AA,
provided by TSFitPy, as a function of S/N for 8\,018 RAVE stars. b)
$\langle$3D$\rangle$ NlTE [O/Fe] uncertainty distribution (solid black
line). We also show the individual distributions of $\langle$3D$\rangle$
NLTE uncertainties due to uncertainties in $\teff$ (blue), log(g) (orange),
and [Fe/H] (green). For comparison, we show the [O/Fe] uncertainty
distribution in 1D LTE (black dashed line). c) $\langle$3D$\rangle$ NLTE
[O/Fe] uncertainties v.s. S/N. d), e) and f) show the $\langle$3D$\rangle$
NLTE [O/Fe] uncertainties as a function of $\teff$, log(g), and [Fe/H],
respectively.}
\label{fig:chi2_snr}
\end{figure*}

\section{Measuring 1D LTE, 1D NLTE, and $\langle$3D$\rangle$ NLTE oxygen abundances in RAVE stars}\label{lte_nlte_abundance_from_rave}


\begin{figure*}
\sidecaption
  \includegraphics[width=12cm]{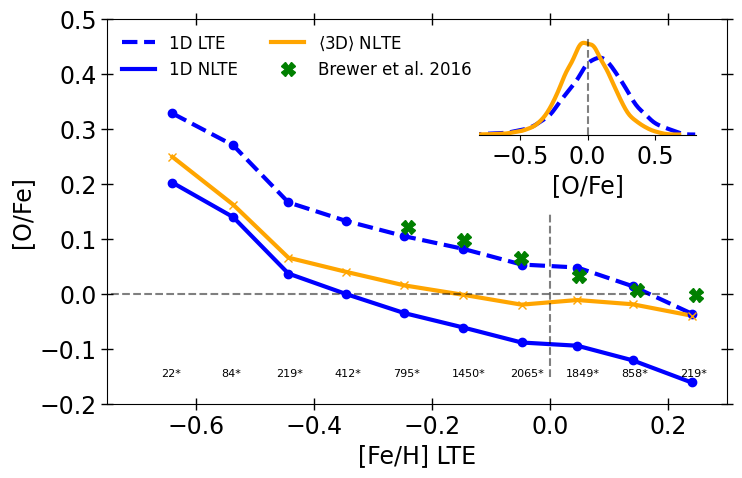}
     \caption{Average [O/Fe] abundances as a function of LTE [Fe/H] ratios for
8\,018 RAVE stars. The blue dots show 1D abundances, in LTE
(blue dashed line) and NLTE (blue solid line). The orange crosses show
$\langle$3D$\rangle$ abundances, in LTE (orange dashed line) and NLTE
(orange solid line). We also show a kernel density distribution of 1D
LTE and $\langle$3D$\rangle$ NLTE [O/Fe] abundances in the top-right
corner. An average trend computed using the 1D LTE data for stars in
the solar neighbourhood from \citet{Brewer2016} is shown with green crosses.}
     \label{fig:OFe_FeH_vs_binned}
\end{figure*}

We adopted the flux-normalized radial-velocity corrected RAVE
spectra\footnote{The RAVE spectra are available at doi:10.17876/rave/
dr.6/019, together with the RAVE radial velocities at doi:10.17876/rave/dr.6/001}
from DR6 \citep{Steinmetz2020b}, used in \citet{guiglion2020}.
Based on the spectral classification
from RAVE DR6, we selected stars classified as normal ('n'), for instance
not suffering from any emission or binarity sign with a probability of
0.99 (see Table 4 of \citealt{Steinmetz2020b}; doi:10.17876/rave/dr.6/004).

To determine oxygen abundances, we need atmospheric parameters $\teff$,
$\logg$, as well as Fe content $\feh$. We adopted the machine-learning
catalog of atmospheric parameters and chemical abundances from
\citet{guiglion2020}. We remind the reader that such machine-learning
catalog was derived using a convolutional neural-network approach (CNN)
trained on high-quality stellar labels from APOGEE, combining RAVE
spectra, \emph{Gaia} DR2 parallaxes, together with 2MASS, WISE and
\emph{Gaia} DR2 magnitudes. The author demonstrated that such a combination
of data allows to break the spectral degeneracies inherent to the RAVE
spectral range, and provided improved labels compared to RAVE DR6 (see
\citealt{guiglion2020} for more details). We selected the labels 
that are within the training set limits, i.e. the most accurate labels 
parametrized by CNN. We selected
RAVE spectra with signal-to-noise ratio above 50 per pixel, for which the
oxygen line is visible enough and atmospheric parameters precise enough
for chemical abundance measurement. The oxygen feature being the strongest
for dwarfs (as demonstrated in Section~\ref{behaviour_oxygen}) , we selected RAVE stars
with $\logg>3.5$, and $6\,300>\teff>5\,700$\,K, corresponding to 8\,018
unique RAVE stars. We notice that this RAVE sample is located within 1\,kpc from the Sun.

Using TSFitPy, we derived LTE and NLTE [O/Fe] ratios in both 1D and
$\langle$3D$\rangle$ for the RAVE stars following the
same method as in the previous section. We adopted a fitting window defined as
[8445.1 - 8448.4]\AA. In total, we were able to measure [O/Fe] ratios in
1D LTE, 1D NLTE, and $\langle$3D$\rangle$ NLTE for 8\,018 RAVE stars. We
show a Kiel diagram of the RAVE stars in \figurename~\ref{fig:Kiel_diagram}.
We show some examples of line fit in \figurename~\ref{spectra_obs}
in the regime 50$<$S/N$<$60 per pixel.
Overall, the algorithm is able to perform very good fitting. We can judge the
quality of the fit by the $\chi^2$ computed between the observation and
the best fit spectrum around the fitting window.

The fitting $\chi^2$ increases for decreasing S/N, as shown in the panel a)
of \figurename~\ref{fig:chi2_snr}. It is expected as the quality of the
fit strongly depend on the noise present in a given RAVE spectrum
\citep{Steinmetz2020b, guiglion2020}. The overall uncertainty of [O/Fe] in
$\langle$3D$\rangle$ NLTE peaks around 0.075, while 1D LTE [O/Fe]
uncertainties peak at 0.082. We can see that the main contributors to
the error budgets are the uncertainties in $\teff$ and log(g). Similarly
to the $\chi^2$, the uncertainty on [O/Fe] decreases with increasing S/N.
However, the error budget slightly increasing for decreasing temperature,
due to increasing blend (panel d). The error is rather constant with log(g),
while it increases for decrease [Fe/H] due to weaker line, and for
super-solar [Fe/H] due to increasing blend.

In Appendix~\ref{appen_corr}, we provide further material regarding the
reliability of our RAVE [O/Fe] ratios computed from full $\langle$3D$\rangle$ NLTE fitting.

\section{Chemical evolution of oxygen with RAVE}\label{chem_evol_rave}

\subsection{1D LTE and NLTE [Fe/O] v.s [Fe/H] diagram}

In \figurename~\ref{fig:OFe_FeH_vs_binned}, we present chemical
abundance trends of 1D-LTE and 1D-NLTE $\ofe$ ratios as a function of 1D-LTE $\feh$,
binned with a step of 0.10 in $\feh$. Both 1D LTE and NLTE abundances
show a monotonic decrease as a function of increasing $\feh$. As suspected
from \figurename~\ref{Sun_syn}, we measure lower $\ofe$ abundances in NLTE
compared to LTE, by about 0.1 dex such a shift being roughly constant with
$\feh$. We note that the dispersion in a given $\feh$ bin is 0.04 dex lower
in 1D-NLTE compared to 1D-LTE, as NLTE tends to reduce the intrinsic dispersion. 

We over-plotted a binned trend of 1D LTE [O/Fe] ratios from \citet{Brewer2016}
with green crosses. We remind the reader that these authors computed 1D LTE
abundances from high-resolution HIRES spectra at Keck Observatory for a
sample of $\sim1\,600$ FGK exo-planet host stars (see \citealt{Brewer2016}
for more details). This sample is mainly confined within 300pc from the Sun.
Our 1D LTE [O/Fe] ratios are in a remarkable agreement with the trend from
\citet{Brewer2016} that the authors derived using the oxygen triplet at
777nm, showing a monotonic decrease
of [O/Fe] with [Fe/H].

\subsection{Effect of $\langle$3D$\rangle$ on [O/Fe] vs. [Fe/H] diagram}

In \figurename~\ref{fig:OFe_FeH_vs_binned}, we present  average trend of
$\langle$3D$\rangle$ NLTE (orange solid line) [O/Fe] ratios. The ratio
decreases for increasing [Fe/H] until solar [Fe/H], then defines a plateau
in the super-solar [Fe/H] regime, contrary to 1D. Interestingly, 1D NLTE and
$\langle$3D$\rangle$ NLTE are rather consistent for [Fe/H]$\lesssim$-0.5, but
then both trends largely deviate at super-solar [Fe/H].

The most interesting comparison to make in this figure is the
difference of behaviour between 1D LTE and $\langle$3D$\rangle$ NLTE.
While both trends decrease towards slightly super-solar [O/Fe] at solar
[Fe/H], $\langle$3D$\rangle$ NLTE [O/Fe] is constant for [Fe/H] with a 
departure of 0.10\,dex. Our $\langle$3D$\rangle$ NLTE results are opposite
to the finding of \citet{Amarsi2019}. Indeed, these authors found a monotonic
decline of $\langle$3D$\rangle$ NLTE oxygen at super-solar [Fe/H] using
the oxygen triplet at 777nm. We notice that \cite{Amarsi2019} computed full 3D NLTE abundance correction rather than $\langle$3D$\rangle$ NLTE spectral fitting.

\begin{figure}
\centering
\includegraphics[width=\columnwidth]{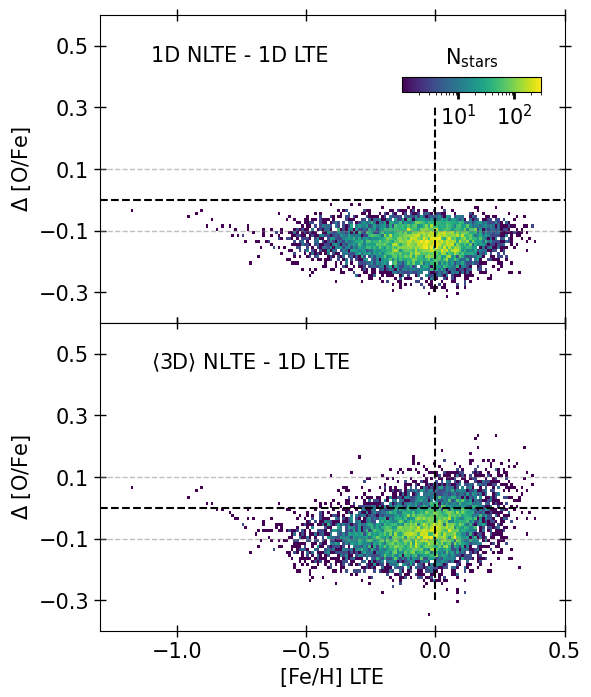}
\caption{2D histogram of the difference in [O/Fe] abundances 1D NLTE - 1D LTE
(left), $\langle$3D$\rangle$ NLTE - $\langle$3D$\rangle$ LTE (middle),
and $\langle$3D$\rangle$ NLE - 1D LTE, as a function of LTE [Fe/H].}
\label{fig:O_departure}
\end{figure} 

In the top panel of \figurename~\ref{fig:O_departure},
we show the difference between 1D NLTE and 1D LTE oxygen
abundances as a function of LTE [Fe/H] in RAVE stars. The
mean difference is equal to -0.14 dex with a mean
dispersion of $0.05$. The correction is rather constant as
a function of [Fe/H], but tends to converge to zero for
[Fe/H]$<$-0.8. In the bottom panel of
\figurename~\ref{fig:O_departure}, consistently with
\figurename~\ref{Sun_syn}, we measure an inversion in
[O/Fe] $\langle$3D$\rangle$ NLTE - 1D LTE corrections,
passing from being negative below solar [Fe/H] to becoming
slightly positive in the super-Solar [Fe/H] regime.

The inversion is somewhat unexpected, because
$\langle$3D$\rangle$ NLTE effects in O I lines are strictly
negative (see Appendix~\ref{appendix_corr}) and they closely match the 1D
NLTE abundance corrections. It is therefore critical to understand
why do we find very small, or even positive
$\langle$3D$\rangle$ - 1D LTE differences for [O/Fe], when
full spectrum synthesis is used in the analysis of abundances
based on the 8\,446\,\AA~lines. As already discussed in
Sect~\ref{syn_3d_nlte}, the effect of Fe I blends is small in
$\langle$3D$\rangle$, but more substantial in 1D LTE. In our
calculations we aim for full internal self-consistency, that
is 1D LTE is used for O and all blends, and the same approach
is adopted in 1D NLTE and $\langle$3D$\rangle$ NLTE. In the
latter two cases, NLTE departure coefficients for the Fe I
blends are used. We have therefore closely inspected the line
fits and the consistency between the best-fit synthetic profiles
in 1D and $\langle$3D$\rangle$. We found that the best data-model
TSFitPy solutions in $\langle$3D$\rangle$ NLTE are typically
characterized by a higher macro-turbulence, compared to the fits
obtained using the 1D hydrostatic model atmospheres
(\figurename~\ref{fig:diff_vmac}, Appendix~\ref{appen_vmac}).
The mean macroturbulence velocity difference between the full
$\langle$3D$\rangle$ NLTE spectrum synthesis and the 1D is on
average $1$ to $2$ kms and it increases slightly with increasing
[Fe/H]. Since increasing macroturbulence corresponds to shallower
profiles, the abundance increases and this effect is the origin
of systematically increasing $\langle$3D$\rangle$ NLTE [O/Fe]
values obtained from the full spectrum synthesis. We emphasize
that the macroturbulence values are not enforced in the calculations:
these values are obtained fully self-consistently with elemental
abundances, which is a robust and well-established procedure
(e.g. \citealt{Bergemann2012, Guiglion2024, Storm2023}). Not surprisingly also
calculations with the averaged 3D models, which lack horizontal
inhomogeneities, require this additional broadening component.
These models only include the depth-dependent proxy of micro-turbulence,
using the recipe adopted in \citet[][see eq. 2]{Semenova2020},
and, thus, they do not represent the full velocity field in the
3D RHD calculations. Indeed, the large-scale velocities are in
particular not accounted for and require an additional broadening
applied during the spectrum synthesis. We therefore consider our
higher $\langle$3D$\rangle$ NLTE values of macroturbulence and
abundances as more realistic and adopt them as the final best
abundances in our work. We note that an in-depth analysis of
velocity power spectra in 3D RHD models in the context of classical
macro-scale velocities would be interesting, but it is beyond
the scope of our work. Such work was performed for red supergiants
\citep{Chiavassa2011}, but a detailed investigation for Sun-like
stars as explored in this work would be warranted and encouraged.

\subsection{Impact of $\langle$3D$\rangle$ NLTE of our understanding of the [O/Fe] vs. [Fe/H] diagram}

How does the inversion in $\langle$3D$\rangle$ NLTE - 1D LTE [O/Fe] ratios
impact our understanding on the chemical evolution of [O/Fe] with [Fe/H],
especially in the super-Solar [Fe/H] regime? The shallow monotonic decline
of [O/Fe] vs. [Fe/H] observed in MW stars goes in favour of a chemical evolution
in the disc where both SN Ia and SN II in a steady rate contribute to the
enrichment of the interstellar medium; as a result it produces a decline in
[O/Fe] with [Fe/H] (see for instance \citealt{Chiappini2003}). Our 1D LTE [O/Fe]
declines with [Fe/H], while our $\langle$3D$\rangle$ NLTE [O/Fe] levels out for
[Fe/H]$>$0. For [Fe/H]$>$, we propose that the flattening of [O/Fe] ratio
could be a sign of the presence of multiple stellar populations in the Solar
neighbourhood.

Indeed, it is now well accepted in the literature that the Solar neighbourhood
(within $\sim2\,$kpc of the Sun) is populated by locally born stars and stars
born in the inner regions of the Milky Way (see for instance
\citealt{casagrande_2011, trevisan_2011, Guiglion2019}). It was proposed that a
significant proportion of stars born in the inner regions migrated towards the
outer regions of the MW disc thanks to radial migration (e.g.
\citealt{roskar_2008, schoenrich_binney_2009, veraciro_2014}). The amplitude of
such a migration depends on the birth radius of the star itself and the age of
the star (see Figure 3. of \citealt{minchev_2013}); to have time to migrate
from the inner-regions to the solar neighbourhood, metal rich stars must be old
(4-8 Gyr). We checked the stellar ages\footnote{We adopted the stellar ages from
RAVE DR6 \citep{Steinmetz2020b} derived by the BDASP pipeline
(Table 9; doi:10.17876/rave/dr.6/008) which is a Bayesian framework demonstrated
in \citet{mcmillan2018}.} distribution of the RAVE stars with [Fe/H]$>$0 used
here: 87\% of the sample is confined between 2 and 8Gyr, with a median value of
$\sim$5Gyr. Star in the super-metallicity regime are not only young, but cover
a wide range of stellar ages, consistently with previous findings (e.g.
\citealt{anders_2017, Dantas2022, Dantas2023}). Recent estimates of stellar birth
radii confirmed that a large fraction of super metal-rich stars come from the
inner regions of the disc (see Figure 8 of \citealt{minchev_2018}).

Galactic chemical evolution models (e.g. \citealt{Matteucci1990, chiappini_2001, minchev_2013})
and observations of very metal-rich stars in the Galactic bulge (e.g.
\citealt{Nepal2023, FeldmeierKrause2022, Rix2024}) support the scenario that the inner
regions of the  Milky Way went through early ($\sim10^7$yr) rapid star
formation. The star-formation rate has been more efficient than in the solar vicinity
by a factor of 10 \citep{Matteucci1990} (contrary to the outer disc with slow and late
star formation, yielding sub-solar [Fe/H] in the outer MW regions; e.g.
\citealt{Bergemann2014GES}). In the super-solar metallicity regime, stars originating
from the inner Milky Way regions (i.e. the bulge) are expected to be super-solar
abundant in $\alphafe$, including oxygen (see for instance Figure 4 of \citealt{Matteucci1990}).

Hence, we propose that the flattening of [O/Fe] at super-solar [Fe/H] observed
in our RAVE data can be interpreted as a mix of stellar populations: one locally
born that exhibit sub-solar [O/Fe] ratios consistently with a pure decline
scenario, and another characterised by high values of [O/Fe] ratios, due to high
star formation efficiency in inner regions of the MW\footnote{This scenario
aligns very well with \citet{Guiglion2019} who suggested that the decrease of
the lithium envelope at [Fe/H] of Solar neighbourhood stars was the result of
mixed stellar population: locally born stars with depleted lithium, together
with radially migrated stars from the inner-regions with high-lithium content.}

\section{Conclusion}\label{conclusiooooonnnn}

In this paper, we focused on constraining the chemical evolution of oxygen in
$\langle$3D$\rangle$ NLTE for the first time with spectra from RAVE. We
summarise here our methodology and main findings:

\begin{itemize}

\item Using the standard spectral fitting code TSFitPy \citealt{Gerber2023, Storm2023}, we performed a careful abundance analysis of the Solar spectrum
at RAVE resolution (Sect.~\ref{lte_oxygen_sun}, \figurename~\ref{Sun_Fit},
\figurename~\ref{Sun_syn}) deriving 1D LTE, 1D NLTE, and $\langle$3D$\rangle$
NLTE [O/Fe] ratios based on the oxygen triplet at 8\,446\,\AA. We found 
$\ofe_{\mathrm{1D~LTE, \bigodot}}=-0.03\pm0.11$, and
$\ofe_{\mathrm{\langle3D\rangle~NLTE, \bigodot}}=-0.01\pm0.09$

\item We selected 8\,018 main-sequence and turn-off stars from the
RAVE survey with high-quality atmospheric parameters from \citet{guiglion2020},
covering $-1<$[Fe/H]$<+0.4$ (see Sect.~\ref{lte_nlte_abundance_from_rave}).
With TSFitPy, we performed high-quality measurements of [O/Fe] ratios in RAVE
spectra (R$\sim$7500; see \figurename~\ref{spectra_obs}, \figurename~\ref{fig:chi2_snr}). 

\item We showed that the difference between 1D NLTE and 1D LTE [O/Fe] ratios
is constant with [Fe/H] in RAVE stars, with a mean difference of -0.14\,dex
(\figurename~\ref{fig:O_departure}). Similar results are obtained when comparing
$\langle$3D$\rangle$ NLTE [O/Fe] to 1D LTE [O/Fe]. When comparing
$\langle$3D$\rangle$ NLTE [O/Fe] to 1D LTE [O/Fe], the difference is negative
below [Fe/H]=0, while it becomes positive for super-solar [Fe/H].

\item The 1D LTE abundance pattern of [O/Fe] vs. [Fe/H] in RAVE stars shows a
monotonic decrease with increasing [Fe/H]. Such a trend matches very well the
1D LTE abundance trend of the very local (d<300pc) exo-planet host stars of
\citet{Brewer2016} based on high-resolution spectra (\figurename~\ref{fig:OFe_FeH_vs_binned}).
The [O/Fe] ratio in $\langle$3D$\rangle$ NLTE decreases with [Fe/H] until
[Fe/H]=0, and then flattens for [Fe/H]>0, contrary to 1D LTE and commonly
measured in literature. Such result is found only when performing full
spectral analysis, contrary to applying $\langle$3D$\rangle$ NLTE corrections.

\item We propose that the flattening of [O/Fe] for [Fe/H]>0 is due to the
interplay between locally born stars (showing negative [O/Fe]) and stars
migrated from the inner MW regions characterised by rapid early star
formation and super-solar [O/Fe].

\end{itemize}

Such work shows that at intermediate resolution, significant information
can be extracted from the oxygen triplet at 8\,446\,\AA. This feature will
be present in low-resolution 4MOST spectra (R$\sim$7\,000 at 8500\,\AA), giving
good insight for oxygen derivation in 4MOST FORMIDABLE-LR survey. In addition, 4MOST low-resolution
spectrograph will cover the oxygen forbidden line at 6\,300\,\AA, as well as
the oxygen triplet at 7\,774\,\AA. The 4MOST high-resolution setup
(R$\sim22\,500$) will also cover the forbidden line at 6\,300\,\AA, allowing
to determine useful abundances. This study demonstrate that that NLTE effects
and realistic treatment of convection using average 3D models are key for
accurate abundance diagnostics even when low- and intermediate-resolution
spectra are used. Future low-resolution
surveys such as 4MOST and WEAVE will have to properly include such effects in their
analysis pipeline.\\

\section{Data availability}\label{data_availability}
The catalogue of 1D LTE, and $\langle$3D$\rangle$ NLTE [O/Fe] and uncertainties is available on the RAVE database at doi:0.17876/rave/dr.6/101). The catalogue is summarized in \tablename~\ref{catalogue}.

\begin{table}
\caption{\label{catalogue}[O/Fe] ratios and uncertainties of the publicly available online catalogue of
$8\,818$ RAVE turn-off and dwarf stars.}
\resizebox{0.49\textwidth}{!}{
\centering
\begin{tabular}[c]{l l l l l}
\hline
\hline
Col & Format & Units & Label            &  Explanations                                      \\
\hline
1   & char   & -     & rave\_obs\_id    & \emph{Gaia} Source ID                              \\
2   & float  & -     & ofe\_1d\_lte     & [O/Fe] ratio computed in 1D LTE                    \\
3   & float  & -     & e\_ofe\_1d\_lte  & uncertainty on [O/Fe]                              \\
4   & float  & -     & ofe\_3d\_nlte    & [O/Fe] ratio computed in $\langle$3D$\rangle$ NLTE \\
4  & float   & -     & e\_ofe\_3d\_nlte & uncertainty on [O/Fe]                              \\
\hline 
\hline
\end{tabular}}
\end{table}

\begin{acknowledgements}
G.G acknowledges the anonymous referee for the
comments and suggestions that improved the readability of the paper. G.G.
sincerely thanks M. Bergemann, N. Storm, and M. Steinmetz for the fruitful
discussions, as well as \^S. Mikolaitis for kindly sharing VUES spectra. G.G.
acknowledges support by Deutsche Forschungs-gemeinschaft (DFG,
German Research Foundation) – project-IDs: eBer-22-59652 (GU 2240/1-1
"Galactic Archaeology with Convolutional Neural-Networks: Realising the
potential of Gaia and 4MOST"). This project has received funding from
the European Research Council (ERC) under the European Union’s Horizon
2020 research and innovation programme (Grant agreement No. 949173). This work
has been partially done in the frame of the Trans-National access program of
Europlanet 2024 RI which has received funding from the European Union's Horizon
2020 research and innovation programme under grant agreement No 871149 project
number 20-EPN2-081.
This work made use of public data from the RAVE survey (https://www.rave-survey.org/).
This work made use of \texttt{overleaf}
\footnote{\url{https://www.overleaf.com/}}, and of the
following \textsc{python} packages (not previously mentioned):
\textsc{matplotlib} \citep{Hunter2007}, \textsc{numpy}
\citep{Harris2020}, \textsc{pandas}
\citep{mckinney-proc-scipy-2010}, \textsc{seaborn}
\citep{Waskom2021}.
\end{acknowledgements}

\bibliographystyle{aa}
\bibliography{cite_r_s}

\clearpage

\appendix

\section{Are full $\langle$3D$\rangle$ NLTE [O/Fe] ratios more precise that
1D LTE [O/Fe] corrected from $\langle$3D$\rangle$ NLTE effects?}\label{appen_corr}

As detailed in the introduction, several past studies derived 1D NLTE (or
$\langle$3D$\rangle$ NLTE) abundances from 1D LTE and applied NLTE (or
$\langle$3D$\rangle$ NLTE) corrections. Throughout the paper, we derived [O/Fe]
ratios from full $\langle$3D$\rangle$ NLTE analysis of RAVE spectra. As an extra
validation, we computed $\langle$3D$\rangle$ NLTE corrections based on the routine
calculate\_nlte\_correction\_line.py from TSFitPy. This routine takes as input the
atmospheric parameters of a given star, and well as its 1D LTE [O/Fe] ratio, and
calculates pure $\langle$3D$\rangle$ NLTE correction without fitting the spectrum. We note that when using abundance corrections, only oxygen abundance is being corrected.

\subsection{Comparison between full $\langle$3D$\rangle$ NLTE and
corrected $\langle$3D$\rangle$ NLTE [O/Fe]}\label{appendix_corr}

\begin{figure}
\centering
\includegraphics[width=\columnwidth]{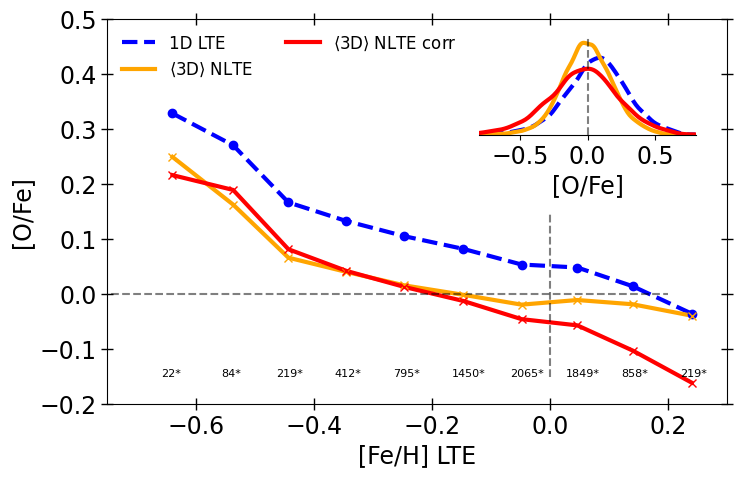}
\caption{Same Figure as \figurename~\ref{fig:OFe_FeH_vs_binned},
together with [O/Fe] ratios computed in RAVE from $\langle$3D$\rangle$
NLTE corrections (in red).}
\label{fig:OFe_FeH_vs_binned_corr}
\end{figure}

\figurename~\ref{fig:OFe_FeH_vs_binned_corr} shows a similar figure as
\figurename~\ref{fig:OFe_FeH_vs_binned}, but we overploted $\langle$3D$\rangle$
NLTE binned [O/Fe] computed from theoretical $\langle$3D$\rangle$ NLTE
corrections. One can see that the corrected $\langle$3D$\rangle$ NLTE [O/Fe]
ratio (in red) monotonically decreases with increasing [Fe/H], similarly to
the 1D LTE and 1D NLTE curves. In the top right corner, the corrected
$\langle$3D$\rangle$ NLTE [O/Fe] distribution (red) is broader ($\sigma=0.26$)
than the 1D LTE ($\sigma=0.22$), and the $\langle$3D$\rangle$ NLTE [O/Fe]
distribution (in orange, $\sigma=0.18$). It suggests that the most precise
method to derive $\langle$3D$\rangle$ NLTE [O/Fe] is by applying full
$\langle$3D$\rangle$ NLTE spectral fitting, as performed throughout
this paper on RAVE data. 

\subsection{[O/Fe] trend with log(g) and $\teff$}

\begin{figure}
\centering
\includegraphics[width=\columnwidth]{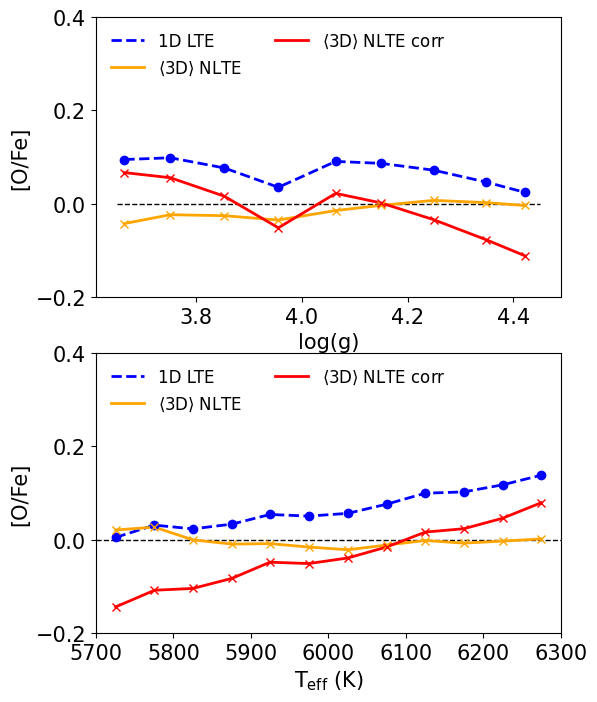}
\caption{[O/Fe] ratios computed in RAVE as a function of log(g) (top) and $\teff$ (bottom).}
\label{fig:abund_vs_teff_logg}
\end{figure}

\figurename~\ref{fig:abund_vs_teff_logg} presents binned [O/Fe] ratios
as a functions of log(g) (top) and $\teff$ (bottom). Both [O/Fe] ratios
computed in 1D LTE (blue, dashed) and corrected from $\langle$3D$\rangle$
NLTE (red, solid) present strong trends with gravity and temperature.
[O/Fe] computed with full $\langle$3D$\rangle$ NLTE fitting presents a
weak trend with log(g) while is constant with temperature, showing a
very good accuracy with log(g) and $\teff$. Several studies already reported
that NLTE abundances computed from corrections may lead to systematics 
(see for instance \citealt{Kirby2018}).

\subsection{Systematic comparison between [O/Fe] derived with O triplets at 777 and 844nm}

In the present paper, we derived for the first time [O/Fe] ratios
in $\langle$3D$\rangle$ NLTE with the O triplet at 844nm. We aim
here at measuring the systematic difference between [O/Fe] computed
with the triplets at 777 and 844\,nm. Both the 777\,nm and the 844nm
lines arise from energy states with similar lower excitation potentials
of approx. 9.14\,eV (777\,nm) and 9.52\,eV (844\,nm), and the upper
energy states are also similar (10.74 vs 10.99\,eV). In addition, we
characterize this difference by using both full $\langle$3D$\rangle$
NLTE fitting and $\langle$3D$\rangle$ NLTE corrected.

To do so, we adopted a sample of 60 FG dwarfs observed by the high
resolution VUES instrument covering the same atmospheric parameter
range as our RAVE sample (\^S. Mikolaitis, private communication).
Such spectra cover the whole optical + near infrared domain, and
have been extensively used for chemical abundance diagnostics (see
\citealt{Stonkute2020, Tautvaisiene2021, Sharma2024} for details on
the instrument and science applications). The VUES spectra were
degraded to RAVE resolution, and we used the 1D LTE atmospheric
parameters from \citet{Tautvaisiene2020, Mikolaitis2019}. We performed
spectral fitting following the same strategy as in
Section~\ref{lte_nlte_abundance_from_rave} of the present paper. We also
computed $\langle$3D$\rangle$ NLTE corrections as in Section~\ref{appen_corr}.
We note that the detailed analysis of the full VUES dataset will be presented
in a subsequent publication (Guiglion et al. in prep.).

\begin{figure}
\centering
\includegraphics[width=\columnwidth]{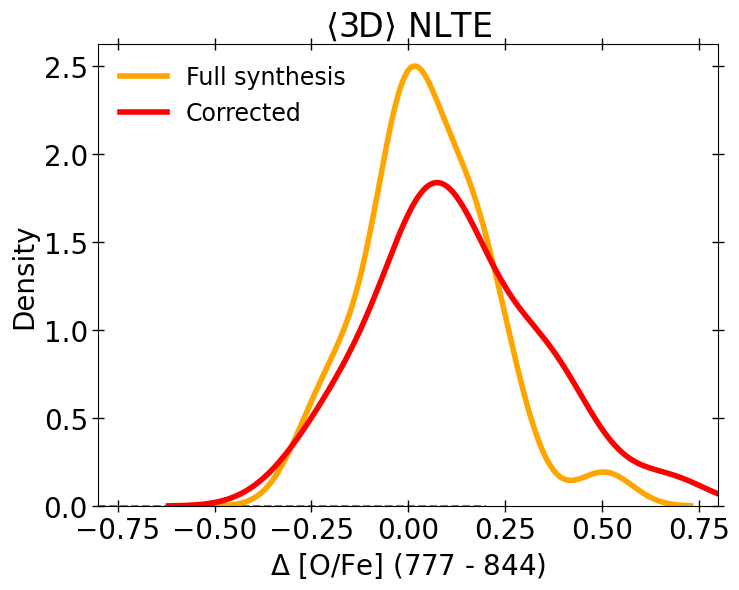}
\caption{Density distribution of the differences between [O/Fe] computed
with the 777nm and 844nm triplets in VUES spectra. The orange curve
corresponds to [O/Fe] ratios computed with full $\langle$3D$\rangle$
NLTE fitting, while the red line corresponds to [O/Fe] ratios computed
from $\langle$3D$\rangle$ NLTE corrections.}
\label{fig:diff_vues}
\end{figure}

In \figurename~\ref{fig:diff_vues}, we presented density distributions
of the differences between [O/Fe] ratios computed with the 777nm and 844nm
O triplets. The orange line, based on [O/Fe] computed with full
$\langle$3D$\rangle$ NLTE fitting is characterized by a mean difference
$\langle\Delta\rangle=+0.04$ dex and a dispersion $\sigma=0.16$. The red line,
based on [O/Fe] computed with $\langle$3D$\rangle$ NLTE corrections is
characterized by $\langle\Delta\rangle=+0.12$ dex, and $\sigma=0.22$ dex.
Here again, in the case of VUES spectra, it is clear that the full
$\langle$3D$\rangle$ NLTE fitting is more precise and accurate than
applying $\langle$3D$\rangle$ NLTE corrections to 1D LTE [O/Fe] ratios.

\section{On macroturbulence velocities}\label{appen_vmac}

\begin{figure}
\centering
\includegraphics[width=\columnwidth]{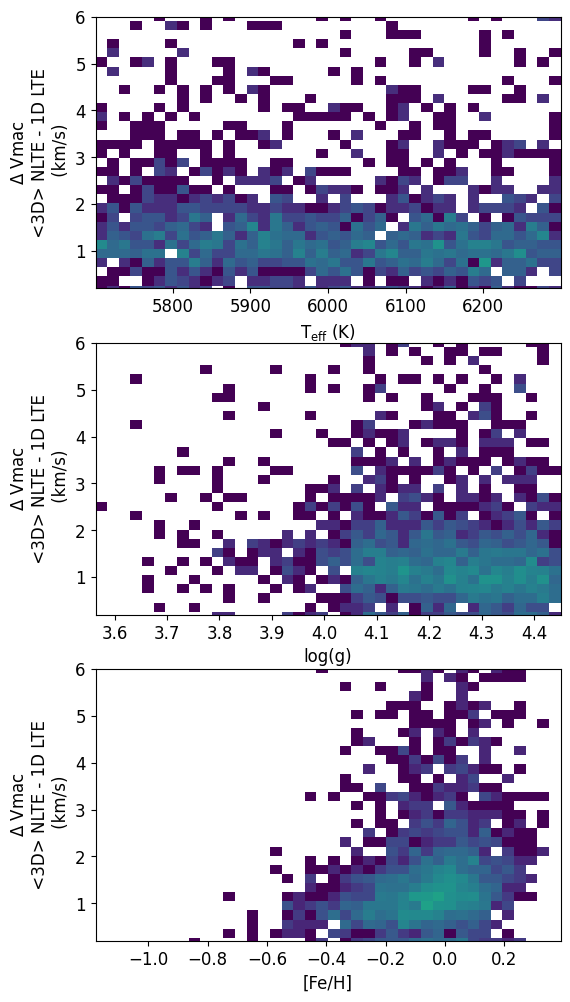}
\caption{Difference in macro-turbulence velocities obtained
during spectral fitting between $\langle$3D$\rangle$ NLTE
and 1D LTE, as a function of $\teff$, log(g), and [Fe/H]
in RAVE stars.}
\label{fig:diff_vmac}
\end{figure}

In \figurename~\ref{fig:diff_vmac}, we present macroturbulence velocity differences 
between $\langle$3D$\rangle$ NLTE and 1D LTE, as a function of input $\teff$, log(g), and [Fe/H]. On average this difference is about 1.7\,\kms. While this difference is constant as a function of $\teff$ and log(g), it increases from 1\,\kms~at [Fe/H]$\sim$-0.5 to 1.7\,\kms~at [Fe/H]$\sim$0.0, and 1.9\,\kms~at [Fe/H]$\sim$+0.1.

\end{document}